\documentclass[notitlepage, amsmath, amssymb, nofootinbib, 11pt, prd, longbibliography]{revtex4-1}

\usepackage[T1]{fontenc}
\usepackage[utf8]{inputenc}
\usepackage{graphicx}
\usepackage{subfig} 

\usepackage{units} 

\usepackage{pifont}

\newcommand{\triD}{\ding{116}} 
\newcommand{\triU}{\ding{115}} 
\newcommand{\circO}{\ding{109}} 

\usepackage{algorithm}
\usepackage{algpseudocode}

\algnewcommand{\IfThen}[2]{\State \algorithmicif\ #1\ \algorithmicthen\ #2}
\algnewcommand{\IfThenElse}[3]{
	\State \algorithmicif\ #1\ \algorithmicthen\ #2\ \algorithmicelse\ #3}

\newcommand{\pqRand}{\textsc{pqRand}}
\newcommand{\ds}{^{}} 
\newcommand{\diff}{\text{d}} 
\newcommand{\bigO}[1]{\mathcal{O}(#1)}
\newcommand{\abs}[1]{\left | #1 \right |}

\newcommand{\sample}[1]{\lbrace #1 \rbrace} 
\newcommand{\dist}{\widehat} 

\newcommand{\can}{\text{E}}
\newcommand{\pqr}{\text{N}}
\newcommand{\Ucan}{U_{\can}\ds}
\newcommand{\Upqr}{U_{\pqr}\ds}

\begin{document}

\title{Reconditioning your quantile function}
\author{Keith Pedersen}
\email{kpeders1@hawk.iit.edu}
\affiliation{Illinois Institute of Technology, Chicago, IL 60616}

\begin{abstract}
Monte Carlo simulation is an important tool for
modeling highly nonlinear systems (like particle colliders and cellular membranes),
and random, floating-point numbers are their~fuel.
These random samples are frequently generated via the inversion method,
which harnesses the mapping of the quantile function~$Q(u)$
(e.g.\ to generate proposal variates for rejection sampling).
Yet the increasingly large sample size of these simulations makes them
vulnerable to a flaw in the inversion method;
$Q(u)$ is \emph{ill-conditioned} in a distribution's tails, stripping precision from its sample.
This flaw stems from limitations in machine arithmetic which are
often overlooked during implementation (e.g.\ in popular C++ and Python libraries). 
This paper introduces a \emph{robust} inversion method, 
which reconditions $Q(u)$ by carefully drawing and using uniform variates.
\pqRand{}, a free C++ and Python package, implements this novel method 
for a number of popular distributions (exponential, normal, gamma, and more).
\end{abstract}

\maketitle{}


\section*{Introduction}

The inversion method samples from a probability distribution~$f$ 
via its quantile function~${Q\equiv F^{-1}}$, 
the inverse of $f$'s cumulative distribution~$F$~\cite{devroye:1986,Press:1992:NRC:148286}.
$Q$ is used to transform a random sample from $U(0,1)$, the uniform distribution over the unit interval,
into a random sample~$\sample{f}$;
\begin{equation}
	\sample{f} 
		= Q\big(\sample{U(0,1)}\big)
	\;.
\end{equation}
This scheme is powerful because quantile functions are formally exact.
But any real-world implementation will be \emph{formally inexact}
because: (i)~A source of true randomness is generally not practical 
(or even desirable), while a repeatable pseudo-random number generator (PRNG) is never perfect. 
(ii)~The uniform variates $u$ and their mapping $Q(u)$ use finite-precision machine arithmetic. 
The first defect has received the lion's share of attention,
leaving the second largely ignored.
As a result, common implementations of inversion sampling lose precision in the tails of~$f$.

This leak must be subtle if no one has patched it.
Nonetheless, the loss of precision commonly exceeds \emph{dozens} of ULP (units in the last place)
in a distribution's tails. Contrast this to library math functions ({\tt sin}, {\tt exp}),
which are painstakingly crafted to deliver no more than \emph{one}~ULP of systematic error.
When the inversion method loses precision, it produces inferior, repetitive samples,
to which Monte Carlo simulations \emph{may} become sensitive
as they grow more complex, drawing ever more random numbers.
Proving that the effect is negligible 
is incredibly difficult, so the best alternative is to use 
the most numerically stable sampling scheme possible with floating point numbers --- 
\emph{if}~it is not too slow.
The robust inversion method proposed here is 80--100\% as fast as the original.

To isolate the loss of precision, we examine the three independent steps of inversion sampling:
\begin{enumerate}
	\setlength\itemsep{0.49ex} 
	\setlength\parskip{-0.5ex}
	\setlength\itemindent{1em} 
	
	\item Generate random bits (i.i.d.\ coin flips) using a PRNG.\label{inv:PRNG}
	
	\item Convert those random bits into a uniform variate $u$ from $U(0,1)$.\label{inv:U}
	
	\item Plug $u$ into $Q(u)$ to sample from the distribution $f$.\label{inv:Q}
	
\end{enumerate}
The first two steps do not depend on $f$, so they are totally generic (a major virtue of the method).
Of them, step~\ref{inv:PRNG} has been exhaustively studied~\cite{Knuth:1997:ACP:270146,
L'Ecuyer:1997:URN:268437.268461, L'Ecuyer:2007:TCL:1268776.1268777}, 
and is essentially a solved problem
--- when in doubt, use the Mersenne twister~\cite{Matsumoto:1998:MTE:272991.272995,L'Ecuyer:2007:TCL:1268776.1268777}.
Step~\ref{inv:Q} has been validated using real analysis~\cite{devroye:1986, steinbrecher_shaw_2008},
so that known quantile functions need only be translated into computer math functions.

This leaves step~\ref{inv:U} which,
at first glance, looks like a trivial coding task to port random bits into a real-valued~$Q$.
Yet computers cannot use real numbers, and neglecting this fact is dangerous 
--- using this as its central maxim,
this paper conducts a careful investigation of the inversion method from step~\ref{inv:U} onward.
Section~\ref{sec:flip-flop} begins by using the condition number to probe step~\ref{inv:Q},
finding that a distribution's quantile function is numerically unstable in its tails.
This provides a sound framework for Sec.~\ref{sec:Upqr} to find the subtle flaw in 
the canonical algorithm for drawing uniform variates (step~\ref{inv:U}).
A \emph{robust} inversion method is introduced to fix both problems, 
and is empirically validated in Sec.~\ref{sec:results}
by comparing the near-perfect sample obtained from the \pqRand{} package 
to the deficient samples obtained from standard C++ and Python tools.


\section{$\boldsymbol{Q}$ are ill-conditioned, but they do not have to be}
\label{sec:flip-flop}

Real numbers are not countable, so computers cannot represent them. 
Machine arithmetic is limited to a countable set like rational numbers $\mathbb{Q}$.
The most versatile rational approximation of $\mathbb{R}$ are 
\emph{floating point} numbers, or ``floats''
--- scientific notation in base-two $(m\times2^E)$.
The precision of floats is limited to $P$,
the number of binary digits in their mantissa $m$, 
which forces relative rounding errors of order ${\epsilon\equiv2^{-P}}$
upon every floating point operation~\cite{Goldberg:1991:CSK:103162.103163}.
The propagation of such errors makes floating point arithmetic formally inexact.
In the worst case, subtle effects like cancellation can 
degrade the \emph{effective} (or de facto) precision 
to just a handful of digits. Using floats with arbitrarily high~$P$ 
mitigates such problems, but is usually emulated in software --- an expensive cure.
Prudence usually restricts calculations to the largest precision widely supported in hardware,
\emph{binary64} ($P=53$), commonly called ``double'' precision.

Limited $P$ makes the intrinsic stability of a computation an important consideration; 
a result should not change dramatically when its input suffers from a pinch of rounding error.
The numerical stability of a function~$g(x)$ can be quantified via its condition number~$C(g)$
--- the relative change in $g(x)$ per the relative change in $x$~\cite{Einarsson:2005:ARS:1088898}
\begin{equation}\label{eq:cond-numb}
	C(g)
		\equiv \abs{\frac{g(x + \delta x) - g(x)}{g(x)} \Big/ \frac{\delta x}{x}}
		=  \abs{x\,\frac{g^\prime(x)}{g(x)}} + \bigO{\delta x}
	\;.
\end{equation}
When an $\bigO{\epsilon}$ rounding error causes $x$ to increment to the next
representable value, $g(x)$ will increment by $C(g)$ representable values.
So when $C(g)$ is large (i.e.\ ${\log_2 C(g)\to P}$),
$g(x)$~is \emph{ill-conditioned} and imprecise;
the tiniest shift in $x$ will cause $g(x)$ to hop over an \emph{enormous} number of values
--- values through which the real-valued function passes, 
and which are representable with floats of precision $P$,
but which cannot be attained via the floating point calculation~$g(x)$.
The condition number should be used to avoid such numerical catastrophes.

\begin{figure}[htb]
\subfloat[]{\includegraphics[scale=1.]{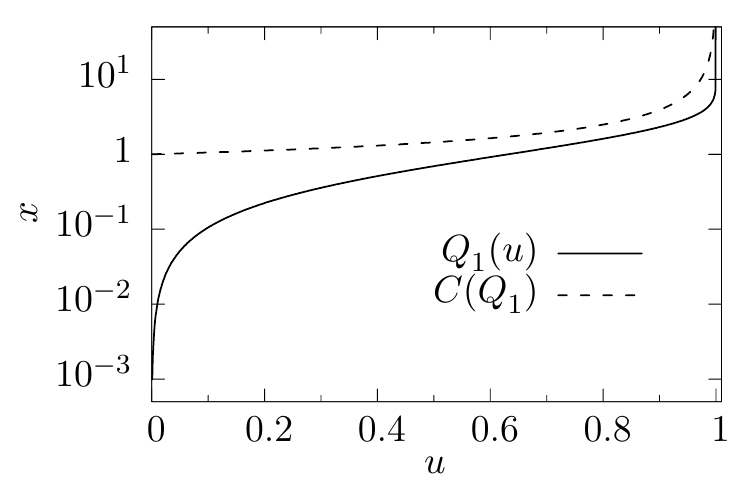}\label{fig:exp-Q1}}
\hspace{2em}
\subfloat[]{\includegraphics[scale=1.]{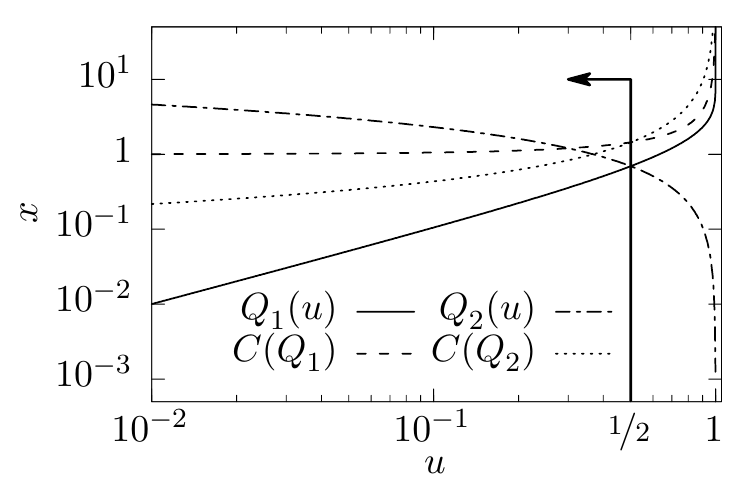}\label{fig:flip-flop}}
\caption{The $\lambda=1$ exponential distribution $f(x)=e^{-x}$;
(a) the quantile function $Q_1\ds$ (solid) and its condition number (dashed) and
(b) the ``quantile flip-flop'' --- in the domain $0<u\le\nicefrac{1}{2}$,\\ 
each $Q$ maps out half of $f$'s sample space while remaining well-conditioned.}
\end{figure}

We now have a tool to uncover possible instability in the inversion method, 
specifically in its quantile function $Q$ (step~\ref{inv:Q}).
As a case study, we can examine the exponential distribution 
(the time between events in a Poisson process with rate~$\lambda$, 
like radioactive decay);\footnote
{\nobreak
	${\tt log1p}(x)$ is an implementation of $\log(1+x)$ which sidesteps
	an unnecessary floating point cancellation~\cite{ISO:2012:III}.
}
\begin{align}\label{eq:exp}
	f(x)=\lambda\,e^{-\lambda x}
	\quad \longrightarrow & \quad
	F(x)=1 - e^{-\lambda x}\;; \\
	Q_1\ds(u) = -\frac{1}{\lambda}\log(1-u) = -\frac{1}{\lambda}{\tt log1p}(-u)
	\quad \longrightarrow & \quad
	C(Q_1\ds) = -\frac{u}{(1-u){\tt log1p}(-u)}\label{eq:exp-Q1}
	\;.
\end{align}
A well-conditioned sample from the exponential distribution will require 
$C(Q_1\ds)\le\bigO{1}$ everywhere, but Fig.~\ref{fig:exp-Q1} clearly reveals that
$C(Q_1\ds)$~(dashed) becomes large as $u\to1$. Why is $Q_1\ds$ ill-conditioned there? 
According to Eq.~\ref{eq:cond-numb}, a function can become ill-conditioned 
when it is \emph{steep}~$(\abs{g^\prime/g}\gg1)$,
and $Q_1\ds$~(solid) is clearly steep at both $u=0$ and $u=1$.
These are $f$'s ``tails'' --- a large range of sample space mapped by a 
thin, low probability slice of the unit interval.
Yet in spite of its steepness, $Q_1\ds$~remains
well-conditioned throughout its small-value tail ($u\to0$) because
floats are denser near the origin
--- reusing the same set of mantissae, but with smaller exponents ---
and a denser set of~$u$ allows a more continuous sampling of a rapidly changing $Q_1\ds(u)$.
This extra density manifests as the singularity-softening factor of~$x$ in Eq.~\ref{eq:cond-numb}.
Unfortunately, the same relief cannot occur as $u\to1$, 
where representable $u$ are not dense enough to accommodate $Q_1\ds$'s massive slope.

Because $Q_1\ds$ is ill-conditioned near $u=1$, 
the large-$x$ portion of its sample $\sample{f}$ will be imprecise;
many large-$x$ floats which should be sampled are skipped-over by $Q_1\ds$.
This problem is not unique to the exponential distribution;
it will occur whenever $f$ has two tails,
because one of those tails will be located near $u=1$.
Luckily, $U(0,1)$ is perfectly symmetric across the unit interval, 
so transforming $u\mapsto1-u$ produces an equally valid quantile function;
\begin{equation}\label{eq:exp-Q2}
	Q_2\ds(u) = -\frac{1}{\lambda}\log(u)
	\quad \longrightarrow\quad
	C(Q_2\ds) = -\frac{1}{\log(u)}\;.
\end{equation}
The virtue of using two valid $Q$'s is evident in Fig.~\ref{fig:flip-flop};
for $u \le \nicefrac{1}{2}$, each version is well-conditioned, 
with $Q_1\ds$ sampling the small-value tail $(x\le\text{median})$ and
$Q_2\ds$ the large-value tail $(x\ge\text{median})$.
Since the pair collectively and stably spans $f$'s entire sample space, 
$f$ can be sampled via the composition method;
for each variate, randomly choose one version of the quantile function
(to avoid a high/low pattern),
then feed that $Q$ a random $u$ from $U(0,\nicefrac{1}{2}\rbrack$.

This ``quantile flip-flop'' --- a randomized, two-$Q$ composition
split at the median --- is a simple, general scheme to 
recondition a quantile function which becomes unstable as $u\to1$.
It is also immediately portable to antithetic variance reduction, 
a useful technique in Monte Carlo integration where,
for every $x=Q(u)$ one also includes 
the opposite choice $x^\prime=Q(u^\prime)$~\cite{kroese2013handbook}.
A common convention is $u^\prime\equiv1-u$, 
which can create a negative covariance $\text{cov}(x,x^\prime)$
that decreases the overall variance of the integral estimate.
Generating antithetic variates with a quantile flip-flop is trivial;
instead of randomly choosing $Q_1\ds$ or $Q_2\ds$ for each variate, 
always use both.


\section{An optimally uniform variate is maximally \emph{uneven}}
\label{sec:Upqr}

The condition number guided the development of the quantile flip-flop,
a rather simple way to stabilize step~\ref{inv:Q} of 
the inversion method during machine implementation.
Our investigation now proceeds to step~\ref{inv:U} --- sampling uniform variates.
While steps~\ref{inv:U}~and~\ref{inv:Q} seem independent, 
we will find that there is an important interplay between them;
a quantile function can be destabilized by sub-optimal uniform variates, 
but it can also wreck itself by mishandling \emph{optimal} uniform variates.

\begin{figure}[htb]
{
\renewcommand{\thesubfigure}{$\can$}
\subfloat[even; $B=P$.]{\includegraphics[scale=1.]{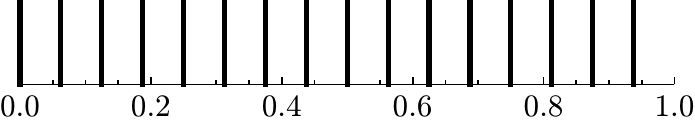}\label{fig:even-comb}}
}
\qquad
{
\renewcommand{\thesubfigure}{$\pqr$}
\subfloat[uneven; $B\to\infty$.]{\includegraphics[scale=1.]{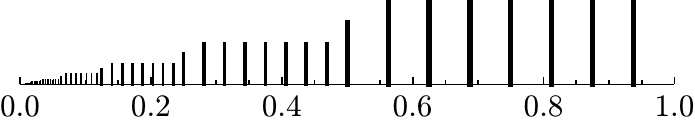}\label{fig:uneven-comb}}
}
\caption{A visual depiction (using floats with $P=4$ for clarity)
of each possible $u$ for (E)~even~$\sample{\Ucan\lbrack0,1)}$ 
and (N)~uneven~$\sample{\Upqr\lbrack0,1)}$.
The height of each tic indicates its relative probability, 
which is proportional to the width of the number-line segment which rounds to it.}
\end{figure}

The canonical method for generating uniform variates is Alg.~\ref{alg:Ucan}~\cite
{Knuth:1997:ACP:270146, 
Press:1992:NRC:148286, 
L'Ecuyer:1997:URN:268437.268461, 
ISO:2012:III, 
python-random, python-random-module, 
numpy-random, numpy-random-kit}; 
an integer is randomly drawn from $\lbrack0,2^B)$, 
then scaled to a float in the half-open unit interval $\lbrack0,1)$.
Using $B\le P$ produces a completely uniform sample space
--- each possible $u$ has the same probability, with a rigidly even spacing of $2^{-B}$ between each.
Using $B=P$ gives the ultimate \emph{even} sample $\sample{\Ucan\lbrack0,1)}$,
as depicted in Fig.~\ref{fig:even-comb} (which uses a ridiculously small $B=P=4$ to aide the eye).
When $B>P$, line~\ref{round-to-float} will be forced to round many large $j$,
as the mantissa of $a$ is not large enough to store every~$j$ with full precision.
As $B\to \infty$, this rounding saturates the floats available in $U\lbrack0,1)$,
creating the \emph{uneven}~$\sample{\Upqr\lbrack0,1)}$ depicted Fig.~\ref{fig:uneven-comb}.
This uneven sample space is still uniform because large~$u$ are more probable,
absorbing more $j$ from rounding (due to their coarser spacing).

\begin{algorithm}[H]
\caption{Canonically draw a random float (with precision $P$) uniformly from $U\lbrack0,1)$}
\label{alg:Ucan}
\begin{algorithmic}[1]
\Require $B\in\mathbb{Z}^+$
	\Comment{$B$ must be a positive integer}
	\State $A\gets\Call{float}{2^B}$
		\Comment{Convert $2^B = (j_\text{max}+1)$ to a float (a power-of-two gets an exact conversion).}
	\Repeat
	\State $j\gets\Call{RNG}{B}$
		\Comment{Draw $B$ random bits and convert them into a integer from $U\lbrack0,2^B)$.}
	\State $a\gets\Call{float}{j}$
		\label{round-to-float}
		\Comment{Convert $j$ to a float with precision $P$. Rounding may occur if $j>2^P$.}
	\Until{$a<A$}
		\label{no1}
		\Comment{If $B>P$ and $j$ rounds to $A$, the algorithm should not return 1.}
	\State\Return $a/A$
\end{algorithmic}
\end{algorithm}

Depending on the choice of~$B$, Alg.~\ref{alg:Ucan} can generate 
uniform variates which are either even or uneven, but which is better?
There seem to be no definitive answers in the literature
--- which is likely why different implementations choose different $B$ ---
so we will have to find our own answer.
We start by choosing the \emph{even} uniform variate $\sample{\Ucan\lbrack0,1)}$
as the null hypothesis, for two obvious reasons: 
(i)~Fig.~\ref{fig:even-comb} certainly \emph{looks} more uniform and 
(ii)~taking $B\to\infty$ does not seem practical.
However, we will soon find that perfect evenness has a subtle side effect
--- it forces all quantile functions to become ill-conditioned as $u\to0$, 
even if they have an excellent condition number!

The condition number implicitly assumes that $\delta x$ is vanishingly small.
This is true enough for a generic float, 
whose $\delta x = \bigO{\epsilon\, x}$ is much small than $x$.
But the even uniform variates have an \emph{absolute} spacing of ${\delta u = \epsilon}$.
To account for a finite $\delta x$, we define a function's \emph{effective} precision
\begin{equation}\label{eq:effective-precision}
	P^*(g) 
		\equiv \abs{\frac{g(x+\delta x) - g(x)}{g(x)}} 
		= \delta x\, \abs{\frac{g^\prime(x)}{g(x)}} + \bigO{\delta x^2}\,.
\end{equation}
Like $C(g)$, a large effective precision $P^*(g)$ indicates an ill-conditioned calculation.
For a generic floating point calculation $\delta x = \bigO{\epsilon\, x}$, 
so $P^*$ reverts back to the condition number (${P^*(g) \approx \epsilon\, C(g)}$).
But feeding even uniform variates into a quantile function gives $\delta u = \epsilon$, so 
\begin{equation}\label{eq:eff-prec-can}
	P^*_\can(Q)
		= \epsilon \abs{\frac{Q^\prime(u)}{Q(u)}} + \bigO{\epsilon^2}
		\,.
\end{equation}
Calculating $P^*_\can(Q)$ for the quantile flip-flop of Fig.~\ref{fig:flip-flop}
indicates that \emph{both} $Q$ become ill-conditioned as $u\to0$ (where $Q$ becomes steep), 
in stark opposition to their excellent condition numbers.
That using even uniform variates will break a quantile flip-flop is 
a problem not unique to the exponential distribution; 
it occurs whenever $f$ has a tail (so that $\abs{Q^\prime/Q}\to\infty$ as $u\to0$).

The reduced effective precision $P^*_\can(Q)$ caused by even uniform variates 
creates sparsely populated tails;
there are many extreme values which $\sample{f}$ will never contain,
and those which it does will be sampled too often.
$\sample{\Ucan\lbrack0,1)}$ is simply \emph{too finite};
$2^P$ even uniform variates can supply no more than $2^P$ unique values.
This implies that the \emph{uneven} sample $\sample{\Upqr\lbrack0,1)}$ will restore quantile stability,
since its denser input space ($\delta u = \bigO{\epsilon u}$)
will stabilize $P^*_\pqr(Q)$ near the origin.
These small~$u$~expand the sample space of~$\sample{f}$ many times over,
making its tails \emph{far less} repetitive.
And since uneven variates correspond to the limit where 
$B\to\infty$ in Alg.~\ref{alg:Ucan}, they are equivalent to 
sampling $U\lbrack0,1-\epsilon)$ from $\mathbb{R}$ and rounding to the nearest float
--- the next best thing to a real-valued input for~$Q$.

The virtue of using uneven uniform variates also follows from information theory.
The Shannon entropy of a sample space~$X$
counts how many bits of information are conveyed by each variate $x$;
\begin{equation}\label{eq:H}
	H(X) \equiv -\sum_i\ds \Pr(x_i\ds)\log_2\ds \Pr(x_i\ds)\,.
\end{equation}
The sample space of the even uniform variates ($B=P$)
has $n=\epsilon^{-1}$ equiprobable members, so 
\begin{equation}
	H_{\can}\ds = -\sum_{i=1}^{n} \epsilon \log_2\ds(\epsilon) = -\log_2\ds \epsilon = P
	\;.
\end{equation}
This makes sense, since each even uniform variate 
originates from a $P$-bit pseudo-random integer.

The sample space of the uneven $\sample{\Upqr\lbrack0,1)}$ contains
every float in $\lbrack0,1)$, 
which is naturally partitioned into sub-domains $\lbrack2^{-k}, 2^{-k+1})$
with common exponent $-k$. Each domain comprises a fraction 
$2^{-k}$ of the unit interval, and the minimum exponent $-K$ 
depends on the floating point type (although $K\gg1$ for \emph{binary32} and \emph{binary64}). 
The uneven entropy is then the sum over sub-domains, 
each of which sums over the $n/2$ equiprobable mantissae\footnote
{\nobreak
	Ignoring the fact that exact powers of 2 are $\nicefrac{3}{4}$ as probable,
	which makes no difference once $P\agt10$.
}
\begin{equation}
	H_{\pqr}\ds = -\sum_{k=1}^{K}\left(\sum_{i=1}^{n/2} 
				2^{-k}(2\,\epsilon)
				\log_2\ds\left(2^{-k}(2\,\epsilon)\right)\right)
		= \sum_{k=1}^{K} 2^{-k}\,(P-1+k)
		\approx P+1 \quad(\text{for}\;K\gg1)
		\;.
\end{equation}
\emph{One more bit} of information than even variates is not a windfall. 
But $H_\can\ds$ and $H_\pqr\ds$ are the entropies of the \emph{bulk} sample $\sample{U\lbrack0,1)}$.
What is the entropy of the tail-sampling sub-space $U\lbrack0,2^{-k})$?

Rejecting all $u\ge2^{-k}$ in the even sample $\sample{\Ucan\lbrack0,1)}$,
we find that smaller $u$ have less information
\begin{equation}\label{eq:Hcan-k}
	H_{\can}\ds(k)=P-k
	\quad(\text{for }u<2^{-k})
	\;.
\end{equation}
This lack of information in even variates is inevitably mapped to the sample $\sample{f}$, 
consistent with the deteriorating effective precision as $u\to0$.
But for \emph{uneven} uniform variates, the sample space is \emph{fractal};
each sub-space looks the same as the whole unit interval, 
so that ${H_{\pqr}\ds(k)=P+1}$ as before! Every~$u$ has maximal information, 
and a high-entropy input should give a high-precision sample.

Both the effective precision $P^*(Q)$ and Shannon entropy $H$ predict
that using even uniform variates will force a 
well-conditioned quantile function to become ill-conditioned, 
precluding a high-precision sample.
Switching to uneven uniform variates will \emph{recondition} it.
But there is an important caveat; uneven variates are \emph{very} delicate.
Subtracting them from one mutates them back into 
\emph{even} variates (with opposite boundary conditions);
\begin{equation}\label{eq:Upqr-delicate}
	1-\sample{\Upqr\lbrack0,1)}\mapsto\sample{\Ucan(0,1\rbrack}
	\,.
\end{equation}
This is floating point \emph{cancellation}.
The subtraction erases any extra density in the uneven sample,
because it maps the very dense region (near zero) to 
a region where floats are intrinsically sparse (near one).
Conversely, the sparse region of the uneven sample (near one) 
has no extra information to convey when it is mapped near zero, and remains sparse. 
This is why $Q_1\ds$ (Eq.~\ref{eq:exp-Q1}) \emph{must} use {\tt log1p}.


\section{Precision: lost and found}\label{sec:results}

In Sec.~\ref{sec:flip-flop} we conditioned 
an intrinsically imprecise quantile function using a two-$Q$ composition.
Then in Sec.~\ref{sec:Upqr} we determined that uneven uniform variates are required
to \emph{keep}~$Q$ well-conditioned. 
These two practices comprise our \emph{robust} inversion method, 
whose technical details we have deliberately left for the Appendix
because we have yet to prove that it makes a material difference.
If indiscreet sampling decimates the precision of $\sample{f}$, 
it should be quite evident in an experiment!

The quality of a real-world sample $\sample{f}$ can be assessed
via its Kullback-Leibler divergence~\cite{Ben-David:2015sia}
\begin{equation}
 D_\text{KL}\ds(\dist{P}||\dist{Q})=
	\sum_i\ds \dist{P}(x_i\ds)\log_2\ds\frac{\dist{P}(x_i\ds)}{\dist{Q}(x_i\ds)}
 \;.
\end{equation}
$D_\text{KL}\ds$ quantifies the \emph{relative} entropy between 
a posterior distribution~$\dist{P}$ and a prior distribution~$\dist{Q}$ (c.f.~Eq.~\ref{eq:H}).
The empirical $\dist{P}$ is based on the count $c_i\ds$ ---
the number of times $x_i\ds$ appears in $\sample{f}$
\begin{equation}\label{eq:P}
	\widehat{P}(x_i\ds) = c_i\ds / N
\end{equation}
(where $N$ is the sample size).
The \emph{ideal} density $\dist{Q}$ is obtained by mapping~$f$ onto 
floats, using the domain of real numbers $(x_{i,L}\ds,x_{i,R}\ds)$ 
that round to each $x_i\ds$;
\begin{equation}\label{eq:Q}
	\dist{Q}(x_i\ds) = 
		\int_{x_{i,L}\ds}^{x_{i,R}\ds}f(x)\,\diff{x}
		= F(x_{i,R}\ds) - F(x_{i,L}\ds)\;.
\end{equation}
$D_\text{KL}\ds$ does not sum terms where $\dist{P}(x_i\ds)=0$ 
(i.e.\ $x_i\ds$ was not drawn), because ${\lim_{x\to0}\ds x\log x = 0}$.

The $D_\text{KL}\ds$ divergence is not a \emph{metric}
because it is not symmetric under exchange of $\dist{P}$~and~$\dist{Q}$~\cite{Ben-David:2015sia}.
And while $D_\text{KL}\ds$ is frequently interpreted as the information 
\emph{gained} when using distribution~$\dist{P}$ instead of $\dist{Q}$, 
this is not true here.
Consider a PRNG which samples from $\dist{Q}=U(0,1)$, but samples so poorly 
that it always outputs $x=0.5$ (and thus emits zero information).
Its $D_\text{KL}\approx P$ is clearly the precision \emph{lost} by $\dist{P}$ (the generator).
In less extreme cases, since $\dist{Q}$ is the most precise distribution possible given
floats of precision $P$, 
any divergence denotes how many bits of precision were \emph{lost}.

Our experiments calculate $D_{KL}\ds$ for samples of the 
$\lambda=1$ exponential distribution generated via the inversion method.
We use GNU's {\tt std::mt19937} for our PRNG ($B=32$), 
fully seeding its state from the computer's environmental noise 
(using GNU's {\tt std::random\_device}).
Calculating $D_\text{KL}$ requires recording the count for each unique float, 
and an accurate $D_\text{KL}$ requires a very large sample size ($N\gg P$, 
so that $\dist{P}\to\dist{Q}$ in the case of perfect agreement).
To keep the experiments both exhaustive and tractable,
and with no loss of generality, we use \emph{binary32} ($P=24$, or single precision).
Since double precision is governed by the same IEEE 754 standard~\cite{P754:2008:ISF}, 
and both types use library math functions with $\bigO{\epsilon}$ errors,
the $D_\text{KL}\ds$ results for \emph{binary64} will be identical.\footnote
{\nobreak
	A \emph{binary64} experiment is tractable, just not exhaustive. 
	Memory constraints require intricate simulation of tiny sub-spaces of 
	the unit interval, to act as a representative sample of the whole.
}

The first implementation we test is GNU's {\tt std::exponential\_distribution}, 
a member of the C++11 {\tt \textless random\textgreater} suite,
which obtains its uniform variates from 
{\tt std::generate\_canonical}~\cite{GNU:random-h, GNU:random-tcc}.
Given our PRNG, these uniform variates are equivalent to 
calling Alg.~\ref{alg:Ucan} with ${B=32}$ and ${P=24}$.
This creates a \emph{partially} uneven sample $\sample{U_\text{P}\ds\lbrack0,1)}$,
with ${B-P=8}$ bits more entropy than even variates.
GNU's implementation feeds these uniform variates into $Q_1\ds$ (Eq.~\ref{eq:exp-Q1}), 
\emph{but without removing its cancellation} by using {\tt log1p}.
As predicted by Eq.~\ref{eq:Upqr-delicate}, the cancellation strips any extra entropy
from the partially uneven variates ($B>P$), converting then into even ones ($B=P$).

\begin{figure}[b]
\includegraphics[scale=1.]{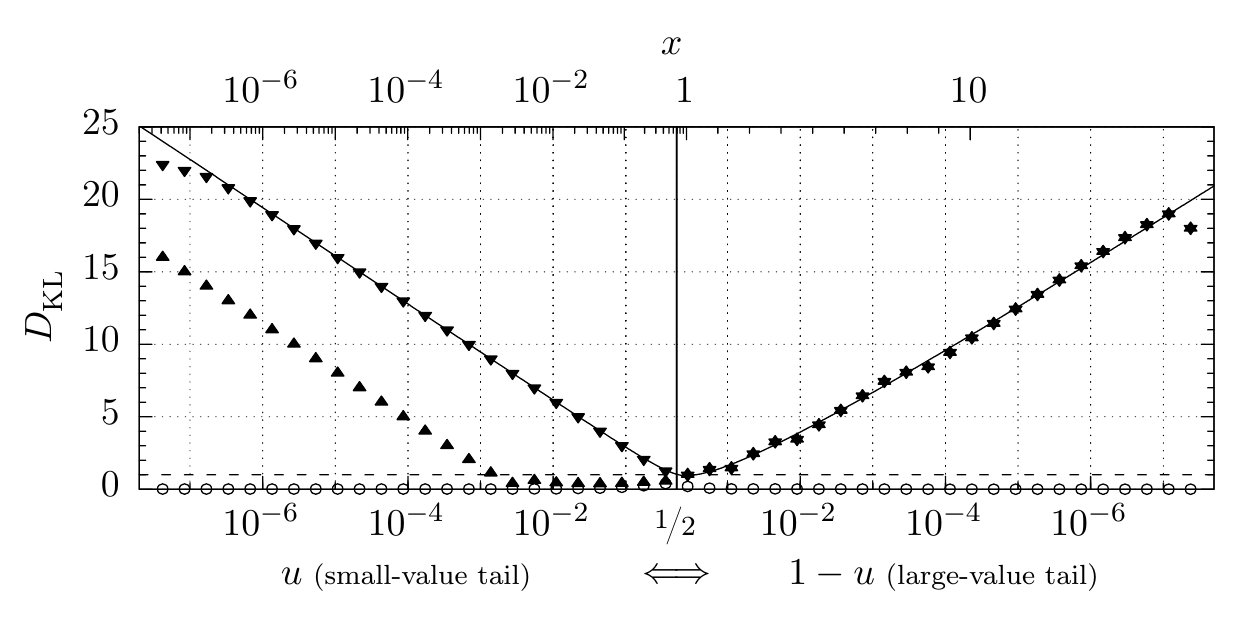}
\caption{The bits of precision \emph{lost} ($D_\text{KL}$) when sampling the 
$\lambda=1$ exponential distribution via
({\scriptsize \triD{}})~GNU's {\tt std::exponential\_distribution}, 
({\scriptsize \triU{}})~GNU's implementation modified to use {\tt log1p}, 
and ({\scriptsize \circO{}})~our robust inversion method (\pqRand{}).
The median $(u=\nicefrac{1}{2})$ bisects the sample-space into two tails,
with improbable values near the left and right edge.
The sampled variate $x=Q(u)$ is shown on the top axis.
Each data point calculates $D_\text{KL}$ for a domain $u\in\lbrack2^{-k},2^{-k+1})$, 
with a sample size of ${N\approx10^9}$ for each point.
The solid line is \emph{not a fit}, but the loss of precision predicted by Eq.~\ref{eq:Hcan-k}
(scaled by $C(Q)$, because precision is lost at a slower pace when $Q'/Q<1$).
The dotted line is a 1-bit threshold.
\label{fig:Dkl}}
\end{figure}

Figure~\ref{fig:Dkl} shows the bits of precision lost by three samples using the same PRNG seed,
with the median at the center and increasingly improbable values near the edges
--- a format which becomes easier to understand by referring to the top axis, 
which shows the $x=Q(u)$ sampled by the various~$u$.
GNU's {\tt std::exponential\_distribution}~({\scriptsize \triD{}})
exhibits a clear and dramatic loss of precision as variates gets farther from the median (and more rare).
This imprecision agrees exactly with 
the prediction of $H_\can\ds(k)$ (Eq.~\ref{eq:Hcan-k}, solid line)
--- even uniform variates have limited information, 
and every time $u$~becomes half as small (so that $x$ is half as probable), 
one more bit of precision is lost.
This loss of precision in the sample is clearly caused by using uniform variates,
which will always happen if $Q_1\ds$ neglects to use {\tt log1p} internally.
Since both Python's {\tt random.expovariate}~\cite{python-random} and 
Numpy's {\tt numpy.random.exponential}~\cite{numpy-random} 
also commit this error, their samples are equally imprecise.

But GNU's {\tt std::exponential\_distribution} could have done better;
it drew \emph{partially uneven} uniform variates ($B=32$, $P=24$), 
then spoiled them via cancellation.
Enabling {\tt log1p} in $Q_1\ds$ and regenerating GNU's sample~({\scriptsize \triU{}})
permits Fig.~\ref{fig:Dkl} to isolate the 
two sources of imprecision identified in Secs.~\ref{sec:flip-flop}~\&~\ref{sec:Upqr}.
(i)~Using {\tt log1p}, $Q_1\ds$ is allowed to be well-conditioned as $u\to0$,
so only the uniform variates themselves can degrade the small-value tail.
Moving left from the median, the \emph{partially} uneven variates
maintain maximal precision until their 8-bit entropy buffer runs dry.
(ii)~Conversely, $Q_1\ds$ is intrinsically ill-conditioned for $u>\nicefrac{1}{2}$
in the large-value tail, so the quality of the uniform variates is irrelevant;
an ill-conditioned quantile function causes an immediate loss of precision.

\pqRand{} generates its sample ({\scriptsize \circO{}})
via our robust inversion method, feeding 
high-entropy, uneven uniform variates $\Upqr(0, \nicefrac{1}{2}\rbrack$ into
a quantile flip-flop which is always well-conditioned 
($Q_1\ds$ samples ${x\le\text{median}}$ and $Q_2\ds$ samples ${x\ge\text{median}}$).
Switching to a quantile flip-flop for this final data series means that,
to the right of the median, the small values shown on 
the bottom horizontal axis are now $u$~instead of $1-u$.
The sample's tails exhibit ideal performance, in stark contrast to the standard inversion method, 
and precision is only lost near the median, 
where the composite $Q$ is a tad unstable ($C(Q)\agt1$, see Fig.~\ref{fig:flip-flop}).
That $D_\text{KL}\ds\approx0$ everywhere, and never exceeds 1 bit, 
is clear evidence that our robust inversion method fulfills its existential purpose,
delivering the best sample possible with floats of precision $P$.
Furthermore, this massive boost in quality arrives at 
${\sim}80/100\%$ the speed of GNU's {\tt std::exponential\_distribution} for \emph{binary32}/\emph{64}
(${\sim}\unit[30/40]{ns}$ per variate on an Intel~i7~@~2.9 GHz with GCC~6.3, optimization~O2).

Similar samples for any rate $\lambda$, as well as many other distributions
(uniform, normal, log-normal, Weibull, logistic, gamma)
are available with \pqRand{}, a free C++ and Python package hosted on GitHub~\cite{pqRand}.
\pqRand{} uses optimized C++ to generate uneven uniform variates (see the Appendix), 
with Cython wrappers for fast scripting.
Yet the usefulness of \pqRand{} is not restricted to 
the rarefied set of distributions with analytic quantile functions;
\pqRand{} uses rejection sampling for its own normal and gamma distributions.
Rejection sampling gives access to \emph{any} distribution~$f(x)$, 
provided that one can more easily sample from the proposal distribution $g(x)\ge f(x)$.
Since the final sample $\sample{f}$ is merely a subset of the proposed sample~$\sample{g}$,
a high-precision~$\sample{f}$ requires a high-precision~$\sample{g}$, 
which can be obtained via our robust inversion method.


\section{Conclusion}

Using the exponential distribution as a case study, we find 
two general sources of imprecision when sampling 
a probability distribution $f$ via the inversion method:
(i)~When $f$ has two tails (two places where ${Q^\prime/Q\gg1}$), 
its quantile function $Q(u)$ becomes ill-conditioned as $u\to1$.
(ii)~Drawing uniform random variates using the canonical algorithm (Alg.~\ref{alg:Ucan})
gives too finite a sample space, making $Q(u)$ ill-conditioned as $u\to0$
(even if $Q(u)$ has a good condition number there).
Both problems can lose dozens of ULP of precision in a sample's tails,
and they are especially problematic for simulations using single precision
--- in the worst case, ${\sim 0.5\%}$ of variates will lose at least a third of their precision.
This vulnerability is found in popular implementations
of the inversion method (e.g.\ GNU's implementation of 
C++11's {\tt \textless random\textgreater} suite~\cite{GNU:random-h},
and the {\tt python.random}~\cite{python-random} and 
{\tt numpy.random}~\cite{numpy-random} modules for Python, and more).

This paper introduces a \emph{robust} inversion method
which reconditions $Q$ by combining
(i) \emph{uneven} uniform variates (Alg.~\ref{alg:Upqr}, see Appendix)
with (ii) a quantile flip-flop (a two-$Q$ composition split at the median).
Our method produces the best sample from $f$ possible with floats of precision $P$, 
and is significantly faster than schemes which ``exactly'' sample distributions to 
arbitrary precision~\cite{knuth1976, Flajolet:2011:BMN:2133036.2133051, Karney:2016:SEN:2888419.2710016}.
The precision of a random sample is especially important for 
large, non-linear Monte Carlo simulations, which can draw 
\emph{so many} numbers that they may be sensitive to this vulnerability.
Since it is generally difficult to exhaustively validate large simulations 
--- in this case, to prove that a loss of precision in the tails has only negligible effects ---
the best strategy is to use the most numerically stable components at 
every step in the simulation chain, provided they are not prohibitively slow.
To this end, we have released \pqRand{}~\cite{pqRand}, a free C++ and Python implementation of our
robust inversion method, which is 80--100\% as fast as standard inversion sampling.


\section{Acknowledgements}

Thanks to Zack Sullivan for his invaluable help and editorial suggestions, 
and to Andrew Webster for lowering the activation energy. This work was supported by the
U.S.\ Department of Energy under award No.~DE-SC0008347
and by Validate Health LLC.


\appendix* 
\section{Drawing uneven uniform variates}

In Sec.~\ref{sec:Upqr} we saw that the best uniform variates are uneven, 
obtained by taking $B\to\infty$ in Alg.~\ref{alg:Ucan}.
Since this will take an infinite amount of time, we must devlop an alternate scheme.
A clue lies in the bitwise representation of the \emph{even} uniform variate from Alg.~\ref{alg:Ucan}, 
for which every $u<2^{-k}$ has a reduced entropy $H_\can\ds = P-k$ (Eq.~\ref{eq:Hcan-k}).
When $B=P$, Alg.~\ref{alg:Ucan} draws an integer $M$ from $\lbrack 0, 2^P)$,
then converts it to floating point.
Inside the resulting float, the mantissa is stored as the integer~$M^*$,
which is just the original integer~$M$ with its bits shifted left until $M^*\ge2^{P-1}$.
This bit-shift ensures that any $u<2^{-k}$ always has at least $k$~trailing zeroes in $M^*$;
zeroes which contain no information.
Filling this always-zero hole with new random bits will restore maximal entropy.

\begin{algorithm}[H]
\caption{Draw an \emph{uneven} random float (with precision $P$) 
uniformly from $U(0,\nicefrac{1}{2}\rbrack$}
\label{alg:Upqr}
\begin{algorithmic}[1]
\Require $B\ge P$
	\State $n\gets1$
		\Comment{We return $j/2^n$. Starting at $n=1$ ensures final scaling into $(0,\nicefrac{1}{2}\rbrack$}.
	\Repeat
		\State $j\gets\Call{RNG}{B}$
			\Comment{Draw $B$ random bits and convert them into a integer from $U\lbrack0,2^B)$.}
		\State $n\gets n + B$
	\Until{$j>0$}\label{find-first-set-bit}
		\Comment{Draw random bits from the infinite stream until we find at least one non-zero bit.}
	\If{$j < 2^{P+1}$}\label{require-P+2-bits}
		\Comment{Require $S\ge P+2$ significant bits.}
	\State $k\gets0$
	\Repeat
		\State $j\gets2 j$
		\State $k\gets k+1$
	\Until{$j \ge 2^{P+1}$}\label{shift-left}
		\Comment{Shift $j$'s bits left until $S=P+2$.}
	\State $j\gets j + \Call{RNG}{k}$
		\Comment{The leftward bit shift created a $k$-bit hole; fill it with $k$ fresh bits of entropy.}
	\State $n\gets n+k$
		\Comment{Ensure that the leftward shift doesn't change $u$'s course location.} 
	\EndIf
	\IfThen{$j\text{ is even}$}{$j\gets j+1$}
		\Comment{Make $j$ odd to force proper rounding.}
	\State \Return $\Call{float}{j}/\Call{float}{2^n}$
		\Comment{Round $j$ to a float using R2N-T2E.}
\end{algorithmic}
\end{algorithm}

Given the domain required by a quantile flip-flop,
Alg.~\ref{alg:Upqr} samples uneven $\sample{\Upqr(0,\nicefrac{1}{2}\rbrack}$
from the half-open, half-unit interval.
It works by taking $B\to\infty$, yet knowing that
floating point arithmetic will truncate the infinite bit-stream to $P$ bits of precision. 
As soon as the RNG returns the first~1 (however many bits that takes),
only the next $P+1$ bits are needed to convert to floating point;
$P$~bits to fill the mantissa, and two extra bits for proper rounding.
To fix $u$'s coarse location, 
the first loop (line~\ref{find-first-set-bit}) finds the first significant bit.
The following conditional (line~\ref{require-P+2-bits}) requires $S\ge P+2$ significant bits.
If $S$ is too small, $j$'s bits are shifted left 
until the most significant (leftmost) bit slides into the $P+2$ position (line~\ref{shift-left}).
Then the vacated space on the right is filled with new random bits,
and the leftward shift is factored into~$n$,
so that only $u$'s fine location changes (enhancing precision while preserving uniformity).
Finally, the integer is rounded into $(0,\nicefrac{1}{2}\rbrack$.\footnote
{\nobreak
	We exclude zero from the output domain of Alg.~\ref{alg:Upqr} because,
	while theoretically possible, it will \emph{never happen} (given a reliable RNG).
	Returning zero in \emph{binary32} (single precision) would require 
	drawing more than 150 all-zero bits in the first loop.
	Given a billion cores drawing $B=32$ every nanosecond, 
	that would take $\unit[\bigO{10^{55}}]{years}$
	(although the first variate with sub-maximal entropy would only take $\unit[\bigO{10^{41}}]{years}$).
	For \emph{binary64}, the numbers get ridiculous.
}

Algorithm~\ref{alg:Upqr} needs two extra bits to maintain uniformity
when $j$ is converted to a float.
With few exceptions, exact conversion of integers larger than~$2^P$ is not possible
because the mantissa lacks the necessary precision.
Truncation $j$ won't work because $j<2^{n-1}$, 
so Alg.~\ref{alg:Upqr} would never return $u=\nicefrac{1}{2}$,
a value needed by a quantile flip-flop to sample the exact median.
Since Alg.~\ref{alg:Upqr} must be able to round $j$ up, 
it uses round-to-nearest, ties-to-even (R2N-T2E).
Being the most numerically stable IEEE~754 rounding mode, 
R2N-T2E is the default choice for most operating systems.

Yet R2N-T2E is slightly problematic because Alg.~\ref{alg:Upqr}
is truncating a theoretically infinite bit stream to finite significance $S$.
There are going to be rounding ties, and when T2E kicks in, 
it will pick \emph{even}~mantissae over odd ones, breaking uniformity. 
To defeat this bias, $j$ is made odd. This creates a systematic tie-breaker,
because an odd $j$ is always closer to only one of the truncated options,
without giving preference to the even option.
This system only fails when $S=P+1$, and only the final bit needs removal.
In this case, $j$ \emph{is} equidistant from the two options,
and T2E kicks in. Adding a random buffer bit (requiring $S\ge P+2$) precludes this failure.

An important property of Alg.~\ref{alg:Upqr} is that $u=\nicefrac{1}{2}$ is
\emph{half as probable} as its neighbor, $u=\frac{1}{2}(1-\epsilon)$.
Imagine dividing the domain $\lbrack\nicefrac{1}{4}, \nicefrac{1}{2}\rbrack$ into 
$2^{P-1}$ bins, with the bin edges depicting the representable~$u$ in that domain.
Uniformly filling the domain with $\mathbb{R}$, 
each $u$ absorbs a full bin of real numbers via rounding  
(a half bin to its left, a half bin to its right).
The only exception is $u=\nicefrac{1}{2}$, 
which can only absorb a half bin from the left, making it half as probable.
But recall that $\sample{\Upqr(0,\nicefrac{1}{2}\rbrack}$ is 
intended for use in a quantile flip-flop
--- a regular quantile function folded in half at the median 
($u=\nicefrac{1}{2}$). Since \emph{both} $Q$ map to the median
when they are fed $u=\nicefrac{1}{2}$, the median will be double-counted
\emph{unless} $u=\nicefrac{1}{2}$ is half as probable.

Not only can Alg.~\ref{alg:Upqr} produce better uniform variates than 
{\tt std::generate\_canonical} (see Fig.~\ref{fig:Dkl}),
it does so at equivalent computational speed
(${\sim}\unit[5]{ns}$ per variate using MT19937 on an Intel i7~@~2.9~GHz).
This is possible because line~\ref{require-P+2-bits} is rarely true 
(${\sim}0.1\%$ when $N=64$ and $P=53$), 
so the code to top-up entropy is rarely needed,
and the main conditional branch is quite predictable.
For most variates, the only extra overhead is verifying that $S\ge P+2$, 
then making $j$ odd, which take no time compared to the RNG and R2N-T2E operations.

\bibliography{ReconditioningYourQuantile}

\end{document}